\documentclass[twocolumn,preprintnumbers,prd,floatfix]{revtex4-1}
\usepackage{graphicx,amsfonts,amsmath,amssymb,epsfig,color, mathrsfs,latexsym,url,wasysym}
\usepackage[hypertex]{hyperref}
\newcommand{\be}{\begin{equation}}
\newcommand{\ee}{\end{equation}}

\newcommand{\ba}{\begin{eqnarray}}
\newcommand{\ea}{\end{eqnarray}}
\newcommand{\mpl}{m_{\rm Pl}}
\newcommand{\lrh}{L_{\rm reheat}}

\newcommand{\trh}{T_{\rm reheat}}
\newcommand{\tn}{T_{\rm synthesis}}
\newcommand{\hinf}{H_{\rm inf}}
\newcommand{\hosc}{H_{\rm osc}}
\newcommand{\hdecay}{H_{\rm decay}}

\newcommand{\hh}{H_{\rm heat}}

\newcommand{\Mev}{{\rm MeV}}
\newcommand{\ev}{{\rm eV}}
\newcommand{\Tev}{{\rm TeV}}
\newcommand{\Gev}{{\rm GeV}}
\newcommand{\n}{N_0}
\newcommand{\lh}{L_{\rm heat}}
\newcommand{\ldec}{L_{\rm decay}}
\newcommand{\Th}{T_{\rm heat}}

\newcommand{\tdec}{T_{\rm decay}}
\newcommand{\hdec}{H_{\rm decay}}
\newcommand{\tf}{T_{\rm freeze}}

\newcommand{\Ne}{N_{\rm e}}
\newcommand{\tdom}{T_{\rm dom}}

\newcommand{\hdom}{H_{\rm dom}}
\newcommand{\ldom}{L_{\rm dom}}

\begin{document}
\title{On Cosmological Implications of Holographic Entropy Bound}
\author{Mahdi Torabian}
\affiliation{International Centre for Theoretical Physics, Strada Costiera 11, 34151, Trieste, Italy}
\begin{abstract}
It is argued that the entropy bound places an upper limit of about 1 TeV on the temperature of the last reheating era
in the early Universe. Implications for inflation and non-standard paradigms for beyond the standard model of hot cosmology are studied. \end{abstract}
\maketitle

\section*{Introduction}
The standard model of hot cosmology is a state of the art description of the Universe back to the time of nucleosynthesis of the light elements. The Hubble expansion of the Universe introduces past particle horizon and confined causally (dis)connected regions. The size of the particle horizon can be expressed in terms of the entropy within a Hubble volume. The horizon problem of the standard model is that the entropy contained within the horizon at any early times was less than today \cite{Guth:1980zm,Kolb:1988aj}. It indicates that the present Hubble volume consists of many causally disconnected Hubble volumes of an earlier time. Nevertheles, the CMB observations indicate that the Universe was smooth to a part in a hundred thousands when photons last scattered. Besides, the nucleosynthesis processes seem to have been identical in different independent regions that comprise the present Hubble volume. The Universe thus had been uniform by the time of the observable nucleosynthesis. 
 
The Universe might have experienced several heating epochs with sizable entropy production. The last (observable) heating phase, after which there was no significant entropy injection and the cosmic expansion has been adiabatic, must have introduced sufficient entropy to account for the observable entropy in the Universe. Equally, the Hubble volume that encompasses all presently observable entropy must fit  within the heating volume in comoving coordinates. 

Entropy has long been subject to theoretical studies since a universal bound on the entropy of a system was proposed \cite{Bekenstein:1993dz}. Motivated by the holographic principle \cite{'tHooft:1993gx,Susskind:1994vu},  Fischler and Susskind \cite{Fischler:1998st} have suggested a version of the entropy bound which is applicable to flat Robertson-Walker geometries (see  \cite{KalyanaRama:1998pk,Easther:1999gk,Bak:1999hd,Kaloper:1999tt,Bousso:1999cb,Bousso:1999xy,Bigatti:1999dp}
 for application and generalizations). It is argued that light-like surfaces must be used to relate entropy and area. The bound applies to the entropy passed through the past-directed light-cone and states that the entropy content does not exceed the area of the particle horizon in Planck units.  

In this note the entropy bound is used to put limits on the temperature of heating epochs in the early Universe. In particular, an upper bound of about 1 TeV is placed on the temperature of the last reheating era. If it is from the inflation decay then the bound greatly constrain inflationary models. The observable reheating could be from the decay of heavy particles after the inflaton decay. In that case the bound constrain the spectrum of paradigms beyond the standard model of particle physics. 

\section*{The Horizon and The Entropy Bound}  
The entropy released in the last reheating era, which accounts for the observable entropy, is given by
\be\label{horizon} g_{\rm last}(T_{\rm last} L_{\rm last})^3 = g_0\left(T_0/H_0\right)^3 \n ,\ee
where $g$ is the number of relativistic degrees of freedom and $\n\equiv (L_0H_0)^3\geq 1$ is the present-day number of Hubble spheres in the Universe (Via measuring the curvature scale of the Universe it is possible to place a lower bound on $\n$ as big as $10^5$ \cite{Scott:2006kga}). The comoving volume that encompasses the observable Universe (the present Hubble sphere scaled back to the time of reheating) lays in the reheating volume of physical size $L_{\rm last}$. The reheating volume contains an entropy which is greater than (or equal to) the observable entropy.

Assuming that the entropy in the reheating volume is predominantly from relativistic species, the entropy bound in flat Robertson-Walker geometry indicates that 
\be\label{entropy-bound} g_{\rm last}(T_{\rm last} L_{\rm last})^3 \lesssim (\mpl L_{\rm last})^2. \ee
Removig $L_{\rm last}$ between \eqref{horizon} and \eqref{entropy-bound} an upper bound on the reheating temperature is found as follows
\be\label{temperature} T_{\rm last} \lesssim \mpl \left(H_0/T_0\right)^{1/2}\n^{-1/6},\ee
(the number of relativistic degrees of freedom is dropped, but can be easily recovered). Moreover, there is a lower bound on the last heating temperature so that the nucleosynthesis processes follow afterwards. It is $T_{\rm last} \gtrsim \tn\sim \Mev$. Substituting numbers ($T_0\sim 10^{-4}~\ev$, $H_0\sim 10^{-33}~\ev$ and $\mpl$ is the reduced Planck mass), one finds the allowed reheating temperature as
\be\label{t-window} 1~\Mev \lesssim T_{\rm last} \lesssim 1~\Tev \times \n^{-1/6}.\ee
The Hubble scale during the last heating era is found by applying the Friedmann  equation ($H_{\rm last} \sim T_{\rm last}^2\mpl^{-1}$)
\be\label{h-window} \tn^2\mpl^{-1} \lesssim H_{\rm last} \lesssim \mpl(H_0/T_0)\n^{-1/3}.\ee
One finds that $10^{-15}\ev \lesssim H_{\rm last} \lesssim 10^{-3}\ev\cdot \n^{-1/3}$.
For a given reheat temperature and number of Hubble spheres, the physical size of the reheating volume can be find through \eqref{horizon}. In fact, using \eqref{t-window} bounds can be placed on the size as follows
\be\label{l-window} \tn^{-1}(T_0/H_0)\n^{1/3}\gtrsim L_{\rm last} \gtrsim \mpl^{-1}(T_0/H_0)^{3/2} \n^{1/2}.\ee 
After substituting numbers it reads $10^{-9}H_0^{-1} \n^{1/3} \gtrsim L_{\rm last} \gtrsim 10^{-15}H_0^{-1}\n^{1/2}$.

\section*{Implications for Beyond the Standard Model of Hot Cosmology}
Inflation is the prime paradigm for the early Universe preceding the standard hot cosmology. It is generally assumed that inflation if followed by a heating phase so that, having stretched the Universe, it solves the horizon problem. The produced entropy has to satisfy \eqref{entropy-bound}. Inflation starts at a Hubble scale $\hinf$ and the size of the smooth patch is $\hinf^{-1}$. Inflation expands this patch by an exponential factor so that at the end of inflation the physical size of the Universe is
\be\label{l-end} L_{\rm end} \sim H_{\rm inf}^{-1}e^{\Ne},\ee
where $\Ne$ is the number of e-folds during inflation. The heating phase can follow immediately or can be delayed.  If it follows promptly then effectively $\hinf\sim\hh$ and the Hubble scale during inflation has to be in the range \eqref{h-window}. It is a very low scale inflation. For a given $\hinf$ the number of e-folds $\Ne$ to solve the horizon problem can be read from \eqref{horizon} as
\be e^{\Ne} \sim  \hinf^{1/2}\mpl^{-1/2} (T_0/H_0)\n^{1/3}.\ee
For the allowed range of Hubble scale \eqref{h-window}, one finds
$21 \lesssim \Ne -\frac{1}{3}\ln\n\lesssim 35$.

If the heating is delayed then the Hubble scale during inflation can be assumed higher. The Universe is matter dominated after the end of inflation until the dawn of heating. Assuming that the heating is from the inflaton decay, when the Hubble scale becomes comparable to the decay rate of the inflaton $\hdec\sim \Gamma_\phi=\beta_\phi m_\phi$ ($m_\phi$ is an effective mass scale in an inflationary model) the inflaton decays and heats up the Universe to a temperature
\be\label{t-heat} \Th \sim \beta_\phi^{1/2}\mpl^{1/2}m_\phi^{1/2}.\ee
If the decay products instantaneously thermalize then $\hh\sim\hdecay$. The entropy bound indicates that $\Th$ has to satisfy \eqref{t-window}. Equivalently, one finds
\be\label{inflaton-mass} 10^{-15}\ev \lesssim \beta_\phi m_\phi \lesssim 10^{-3}\ev\times \n^{-1/3}.\ee
If one identifies $m_\phi$ as the inflaton mass and assumes that it is solely gravitationally coupled, namely $\beta_\phi\sim m_\phi^2\mpl^{-2}$, then it should be in the following range
\be\label{mass-range} 10^4 ~\Gev \lesssim m_\phi \lesssim 10^8~\Gev \times \n^{-1/9}.\ee
If the inflaton has a stronger coupling on top of gravitational, then the parameter $\beta_\phi$ is bigger and from \eqref{inflaton-mass} the inflaton is lighter than the exclusively gravitationally coupled one. 

The dimensionless power spectra of scalar and tensor perturbations are respectively given by 
\ba \label{scalar-fluc}\Delta_s^2 &\sim& 2\hinf^{2}(4\pi\mpl)^{-2}\epsilon^{-1} \sim 10^{-9},\\
\label{tensor-fluc}\Delta_t^2 &\sim& 32\hinf^2(4\pi\mpl)^{-2} = r\cdot \Delta_s^2,\ea
where $r<0.2$ from non-observation of tensor modes. The scalar and tensor spectral index respectively are
\be\label{tilts} 1-n_s \sim 6\epsilon-2\eta \approx 0.03,\quad n_t \sim -2\epsilon.\ee

The size of the Universe at the time of heating is
\be\label{l-heat} \lh \sim L_{\rm end} \hinf^{2/3}\mpl^{2/3}\Th^{-4/3}.\ee
For an inflationary scenario to solve the horizon problem, generate observable scalar perturbations, trigger nucleosynthesis and be consistent with the entropy bound, using \eqref{horizon}, \eqref{t-window}, \eqref{l-end}, \eqref{scalar-fluc} and \eqref{l-heat} one finds
\be \label{inflation-parameters} \tn^{1/3}\mpl^{-1/3}(T_0/H_0)\Delta_s \n^{1/3}\lesssim e^{\Ne} \epsilon^{-1/2}\lesssim (T_0/H_0)^{5/6}\Delta_s\n^{5/18},\ee
or approximately $41 \lesssim \Ne - \frac{1}{2}\ln\epsilon - \frac{1}{3}\ln\n\lesssim 46$.

A model of inflation with a given set of parameters ($\hinf, \eta, \epsilon, \beta_\phi, m_\phi, \Ne$) must thus satisfy  \eqref{inflaton-mass}, \eqref{tensor-fluc}, \eqref{tilts} and  \eqref{inflation-parameters}. Models  have to be studies case by case. An extensive study is postponed to a future work.

Moreover, long-lived unstable particles are ubiquitous in paradigms beyond the standard model of particle physics.
If they take over dynamics of the Universe before decaying then the entropy release after the decay dominates over any preexisting entropy. In fact, the observed entropy (in the CMB photons) is originated form the latter decay. It solves the horizon problem and sets off the nucleosynthesis processes. Nevertheless, some period of inflation is needed to account for the size problem of the observable Universe, so that the horizon problem is solved by some late entropy production ({\it i.e.} inflation solely expands the physical size of the Universe but does not solve the horizon problem). The physical size of the Universe is orders of magnitude greater than the  horizon at the time of latter entropy release. The same laws of physics apply everywhere with almost the same initial conditions in all Hubble spheres contained in the Universe. Matter species uniformly decay everywhere and reheat the Universe in a smooth way. 

Matter component $\chi$ is produced with number density $n_\chi^{\rm freeze}$ after the inflaton decay. The number density to the entropy density, $Y_\chi=n_\chi^{\rm freeze}s_{\rm freeze}^{-1}$, is constant in time. It takes over the dynamics at the Hubble scale $\hdom (\sim \tdom^2\mpl^{-1})$ at temperature
\be\label{t-dom} \tdom \sim m_\chi Y_\chi^{\rm freeze}.\ee
Later when the the scale becomes comparable to its defy rate, $\hdec\sim\Gamma_\chi=\beta_\chi m_\chi$, it decays. One notes that
\be\label{beta} \beta_\chi \gtrsim m_\chi^2\mpl^{-2},\ee
where equality applies for gravitational coupling.
The decay products thermalize and reheat the Universe to a temperature given as follows
\be\label{t-reheat} \trh \sim \beta_\chi^{1/2} \mpl^{1/2}m_\chi^{1/2}.\ee
It is subject to the constrain \eqref{t-window}. 
The temperature of the pre-existing plasma from former heating, which has gone through a matter era, drops to 
\be \tdec \sim \beta_\chi^{2/3}m_\chi^{2/3}\mpl^{2/3}\tdom^{-1/3}.\ee
It is (much) smaller than the latter reheat temperature $\trh$. In fact, one finds
\be\label{temperature-relation} \Th \geq \tf \geq \tdom \geq \trh\gg \tdec.\ee
From above and for later use one introduces two parameters $\zeta =\tdom\Th^{-1}\leq 1$ and $\gamma=\trh\tdom^{-1}\leq 1$. The above equation in particular implies
\be\label{T-hierarchy} \beta_\phi^{1/2} m_\phi^{1/2}\mpl^{1/2} \geq m_\chi Y^{\rm freeze}_\chi \geq \beta_\chi^{1/2} m_\chi^{1/2} \mpl^{1/2}.\ee
One has to check if there exists regions  
the parameter space so that equations \eqref{t-window}, \eqref{beta} and \eqref{T-hierarchy} are satisfied and the latter reheat process takes place. 

Energy conservation in matter dominated Universe indicates that the physical size of the Universe at the time of reheating is (assuming $\ldec\sim\lrh$)
\be\label{energy-conservation} \lrh \sim \ldom \hdom^{2/3}\Gamma_\chi^{-2/3} \sim \ldom  \tdom^{4/3}\trh^{-4/3}.\ee
Furthermore, the entropy is constant in comovig volume
\be\label{entropy-conservation} \lh\Th \sim \ldom\tdom\sim\lrh\tdec.\ee
From \eqref{horizon}, \eqref{entropy-bound}, \eqref{energy-conservation} and \eqref{entropy-conservation} one finds that the entropy bound places an upper bound on $\Th$ as follows
\be\label{t-heat-former} \Th \lesssim \mpl^{3/2} (H_0/T_0)^{3/4}\trh^{-1/2}\ \zeta\,\n^{-1/4}.\ee
After applying \eqref{t-window} the bound can be read as
\be\label{t-heat-former-numbers}  \Th \lesssim 10^6\Gev\times\zeta\, \n^{-1/4} (\trh/\Mev)^{-1/2}.\ee 
Furtheremore, from \eqref{horizon} the size of the Universe at the time of heating and reheating can be determined. 
In particular, one finds
\be\label{the-equation} e^{\Ne} \hinf^{-1/3}\Th^{-1/3}\tdom^{1/3}\trh^{-1/3} \sim \mpl^{-2/3}(T_0/H_0)\n^{1/3}.\ee
In the following, two unstable particle species are studied; modulus condensates and thermally/non-thermally produced fermions. 
\paragraph{Scalar Condensates} 
When the Hubble scale becomes of order the mass of scalar field $\varphi$, $\hosc\sim m_\varphi$, it performs coherent oscillations about its minimum and produces pressureless condensates with number density $n_\varphi^{\rm freeze}$. If the modulus mass $m_\varphi$ is less than the inflaton decay rate, $m_\varphi\leq \Gamma_\phi$, then the modulus is released when the the inflaton has already decayed and the Universe is radiation dominated. However, if the mass of the modulus is greater than the decay rate of the inflaton, $m_\varphi\leq\Gamma_\phi$, then it oscillates before the inflaton decays. The Universe is thus filled with the inflaton and the modulus condensates. When the inflaton decays, it heats up the Universe to $\Th$ and produces a huge amount of entropy which dilutes away any preexisting species. Nevertheless, the modulus resumes oscillations and condensates get reproduced. In either case, the number density of the modulus condensates to the entropy density is
\be\label{Y-modulus} Y_\varphi^{\rm freeze} \sim \mpl^{1/2}m_\varphi^{-1}{\rm min}(\Gamma_\phi,m_\varphi)^{1/2}\delta_\varphi^2,\ee
where $\delta_\varphi=(\varphi_{\rm init}/\mpl)$. Eventually, when the Hubble scale becomes comparable to the modulus decay rate the modulus decays and reheats the Universe to $\trh$ given in \eqref{t-reheat}. The parameters $\zeta$ and $\gamma$ are read as follows
\be \zeta \sim \gamma^{-1} \sim m_\phi^{-1/2}{\rm min}(\Gamma_\phi,m_\varphi)^{1/2}\beta_\phi^{-1/2}\delta_\varphi^2.\ee
If the modulus is gravitationally coupled then its mass must be in the range \eqref{mass-range} and the above analysis indicates that it starts oscillation before the inflaton decay ({\it i.e.} $\Gamma_\phi<m_\varphi$ and thus $\zeta\sim1$). One can read the upper bound on $\Th$ from \eqref{t-heat-former-numbers}. If the Universe has always been matter dominated before the last reheating era then the size of the Universe after the end of inflation is
\be L_{\rm end} \sim \hinf ^{-2/3}\mpl^{-2/3} \trh^{1/3}(T_0/H_0)\n^{1/3},\ee
so that the horizon problem is solved by the last reheating phase.
It applies to situations when there are heavy moduli that role the dynamics of the early Universe, like theories with towers of Kaluza-Klein modes. Heavy fields oscillate and decay when the inflaton has not yet decayed so the decay products dilutes away by the infaton decay. Later on, lighter fields in the allowed range \eqref{mass-range} immediately take over, subsequently decay and reheat the Universe. Thus, if there are many heavy fields in the spectrum, there must be at least one in the allowed range \eqref{mass-range} so that when it decays it accounts for the observable entropy. Further studies of modulus dynamics is postponed to  a future work \cite{Torabian:2012}.

\paragraph{Frozen Fermions} 
When the inflaton decays, massive weakly interacting particle species $\psi$ can be found in decay products. 
If $\Th$ is greater than $m_\psi$, $\psi$ particles are relativistic at production and they thermalize in the plasma. When the temperature of the plasma drops below their mass, they become non-relativistic, rapidly annihilate, their number density exponentially decreases and the entropy in them transferred to the relativistic species. When the annihilation rate becomes less than the Hubble scale, $\Gamma^{\rm ann}_\psi\sim n_\psi\langle\sigma^{\rm ann}_\psi v\rangle \lesssim H_{\rm freeze}$, they freeze out and the number density in comoving volume remains constant
\be Y_\psi^{\rm freeze} \sim \alpha_\psi^{-1}\tf^{-1}\mpl^{-1}m_\psi^2.\ee 
The annihilation cross section is parametrized as
$\langle\sigma^{\rm ann}_\psi v\rangle = \alpha_\psi m_\psi^{-2}$
and the freeze out temperature is found as follows
\be m_\psi\sim \tf\ln\big(0.04 \alpha_\psi \mpl m_\psi^{-1}\big).\ee
The frozen-out species take over at a temperature given in \eqref{t-dom}.  
When the Hubble scale becomes comparable to the decay rate, $H_{\rm decay} \sim \Gamma_\psi^{\rm decay} =\beta_\psi m_\psi$,
the $\psi$ particle decay and reheat the Universe to a temperature given in \eqref{t-reheat}. However, since the mass assumed to be  below the heating temperature, one finds that it impossible for generic values of parameters \eqref{T-hierarchy} is satisfied. Therefore, the decay of thermal left overs cannot explain the observable entropy. 

Nevertheless, $\psi$ particles can be produced so that they are too heavy to thermalize ($\Th \lesssim m_\psi<m_\phi$) \cite{Moroi:1999zb}.  
They annihilates until the annihilation rate becomes smaller than the Hubble scale, $\Gamma_\psi^{\rm ann}\sim \hh$. It happens when the number density reaches the critical values $n_\psi^{\rm freeze} \langle\sigma^{\rm ann}_\psi v\rangle \sim \Gamma_\phi^{\rm decay}$. 
Afterward, the number density per comoving volume remains constant
\be Y_\psi^{\rm freeze} \sim \alpha_\psi^{-1}\beta_\phi^{-1/2} m_\phi^{-1/2} m_\psi^2 \mpl^{-3/2}.\ee
The frozen-out species take over the energy density at a temperature given in \eqref{t-dom}. When the Hubble scale becomes comparable to the decay rate, $\psi$ particle decay and reheat the Universe to a temperature given in \eqref{t-reheat}. The parameters $\zeta$ and $\gamma$ are read as follows
\ba \zeta &\sim&  m_\psi^3m_\phi^{-1}\mpl^{-2}\alpha_\psi^{-1}\beta_\phi^{-1},\\ 
\gamma &\sim& m_\psi^{-5/2}m_\phi^{1/2}\mpl^2 \alpha_\psi\beta_\psi^{1/2}\beta_\phi^{1/2}.\ea 
For generic values of parameters there exists a region in the parameter space so that equations \eqref{t-window}, \eqref{beta} and \eqref{T-hierarchy} are simultaneously satisfied. Thus, heavy unstable fermions can take over the dynamics after inflaton decay and reheat the Universe in the allowed range and explain the observable entropy. Further studies of the parameter space is postponed to a future work \cite{Torabian:2012}.

\section*{Conclusions}
In this note an upper bound of about 1 TeV is placed on the temperature of the last reheating era. The last reheating accounts for the observable entropy in the Universe and sets off the nucleosynthesis processes. It can be either from the inflaton decay or from the decay of other  unstable dominating species. If it is from the inflaton decay, the parameters of an inflationary scenario are greatly constrained. The last reheating phase can also be deriven by the decay of unstable particles in the spectrum. The upper bound now applies on the temperature of the latter reheating era and the upper bound on the former heating era can be raised. A limited region in the parameter space allows such a scenario to take place.
Any paradigm for the early Universe cosmology preceding the nucleosynthesis has to satisfy certain criteria which are based on the entropy bound and the solution to the horizon problem. 

Moreover, the closure of the temperature window places an upper bound on the number of Hubble spheres in the Universe $\n\lesssim 10^{36}$. It consequently puts an upper bound on the physical size of the present Universe $H_0^{-1}\lesssim L_0 \lesssim 10^{12} H_0^{-1}$. If there were no cosmological constant then all the present Hubble spheres would have form a single Hubble sphere in the far future and a patient observer could have measured all the entropy produced in the last reheating phase. However, the Universe is approaching an asymptotically de Sitter space and the presence of a cosmological event horizon implies that the unobserved portion of entropy will be concealed for ever. Further studies to investigate a likely link between the screened entropy and the horizon entropy (and thus the value of the cosmological constant) is of great significance.

\acknowledgments
I am grateful to B. S. Acharya,  J. Conlon, P. Creminelli, F. Quevedo, S. Randjbar-Daemi, V. Rubakov, L. Susskind, C. Vafa for discussions/correspondence and to M. M. Sheikh-Jabbari for comments on the inflation part. I am thankful to the ENS Foundation for financial support and to the Isaac Newton Institute for hospitality during the completion of this work.     

\ \newline e-mail: {\tt mahdi@ictp.it}

\end{document}